\documentclass[sigconf]{acmart}
\AtBeginDocument{%
  }

\acmISBN{979-8-4007-2292-9/2026/02}

\usepackage{enumitem}
\usepackage{multirow}
\usepackage{colortbl}
\usepackage{xspace}
\usepackage{amssymb}
\usepackage{varwidth}
\definecolor{bgcolor}{RGB}{242, 242, 242}
\usepackage{makecell}

\begin{document}

\title{ANCHOR: Agentic Noise Creation Framework for Human Simulation and Denoising Recommendation}

 
\author{Xiangming Li}
\affiliation{%
  \institution{Xidian University}
  \department{School of Computer Science and Technology}
  \city{Xi'an}
  \state{Shaanxi}
  \country{China}
}
\email{xiangmli@stu.xidian.edu.cn}

\author{Hua Chu}
\affiliation{%
  \institution{Xidian University}
  \department{School of Computer Science and Technology}
  \city{Xi'an}
  \state{Shaanxi}
  \country{China}
}
\email{hchu@mail.xidian.edu.cn}

\author{Jiaming Liang}
\affiliation{%
  \institution{Xidian University}
  \department{School of Computer Science and Technology}
  \city{Xi'an}
  \state{Shaanxi}
  \country{China}
}
\email{jiamliang@stu.xidian.edu.cn}

\author{Jianan Li}
\affiliation{%
  \institution{Xidian University}
  \department{School of Computer Science and Technology}
  \city{Xi'an}
  \state{Shaanxi}
  \country{China}
}
\email{lijianan@xidian.edu.cn}

\author{Yangtao Zhou}
\authornote{Corresponding Author.}
\affiliation{%
  \institution{Xidian University}
  \department{School of Computer Science and Technology}
  \city{Xi'an}
  \state{Shaanxi}
  \country{China}
}
\email{zhou_yt@stu.xidian.edu.cn}

\author{Qingshan Li}
\affiliation{%
  \institution{Xidian University}
  \department{School of Computer Science and Technology}
  \city{Xi'an}
  \state{Shaanxi}
  \country{China}
}
\email{qshli@mail.xidian.edu.cn}

\author{Wanqiang Yang}
\affiliation{%
  \institution{Shanghai Fairyland Software Co.,Ltd.}
  \city{Shanghai}
  \country{China}}
\email{yangwan@fulan.com.cn}

\renewcommand{\shortauthors}{Xiangming Li et al.}

\begin{abstract}
Distilling accurate user preferences from noisy implicit feedback remains a fundamental bottleneck in recommendation systems, underscoring the critical necessity of recommendation denoising. However, the absence of explicit noise annotations in real-world data prevents models from directly learning realistic noise patterns, forcing existing methods to operate in an unsupervised manner by relying on side information or handcrafted heuristic assumptions. Unfortunately, these methods often suffer from high external costs, limited generalization, or unreliable priors, leading to systematic noise misidentification and the corruption of true user preference representations. To overcome these limitations, we propose a paradigm-level reformulation of recommendation denoising. Rather than indirectly inferring noisy interactions through unsupervised heuristics or handcrafted rules, our Creation-Recognition paradigm proactively creates labeled noisy interactions and trains a dedicated recognizer to identify them. This reformulation transforms recommendation denoising from a heuristic filtering problem into a supervised learning task. Instantiating this paradigm, we present ANCHOR, an agent-based framework inspired by recent "LLM-as-User" research. By proactively simulating user behaviors to generate realistic noise labels, ANCHOR facilitates supervised denoising through a dual-stage process involving noise creation and noise recognition. In the noise creation stage, we simulate realistic user-system interactions by devising a recommender-in-the-loop agentic architecture, which synthesizes diverse out-of-preference noise as well as informative boundary-adjacent noise. For out-of-preference noise, we implement five representative and extensible simulation mechanisms to model common non-preference-driven behaviors, including misclick, curiosity-driven, caption-biased, popularity-biased, and position-biased interactions, thereby approximating major sources of noisy implicit feedback. For boundary-adjacent noise, we introduce an adversarial boundary refinement mechanism where a noise creator synthesizes ambiguous interactions to challenge a noise recognizer, thereby driving the production of informative samples targeting the decision boundary. In the noise recognition stage, leveraging the generated noise labels, we propose a reusable parametric recognizer that integrates collaborative signals and semantic representations to identify noise patterns in real interaction data. This effectively transforms recommendation denoising from an unsupervised challenge into a fully supervised task. Extensive experiments on three benchmarks demonstrate that ANCHOR serves as a robust, model-agnostic solution that significantly outperforms state-of-the-art recommendation denoising methods. 
\end{abstract}




\maketitle

\begin{figure}[t]
\centering
\includegraphics[width=1.0\linewidth]{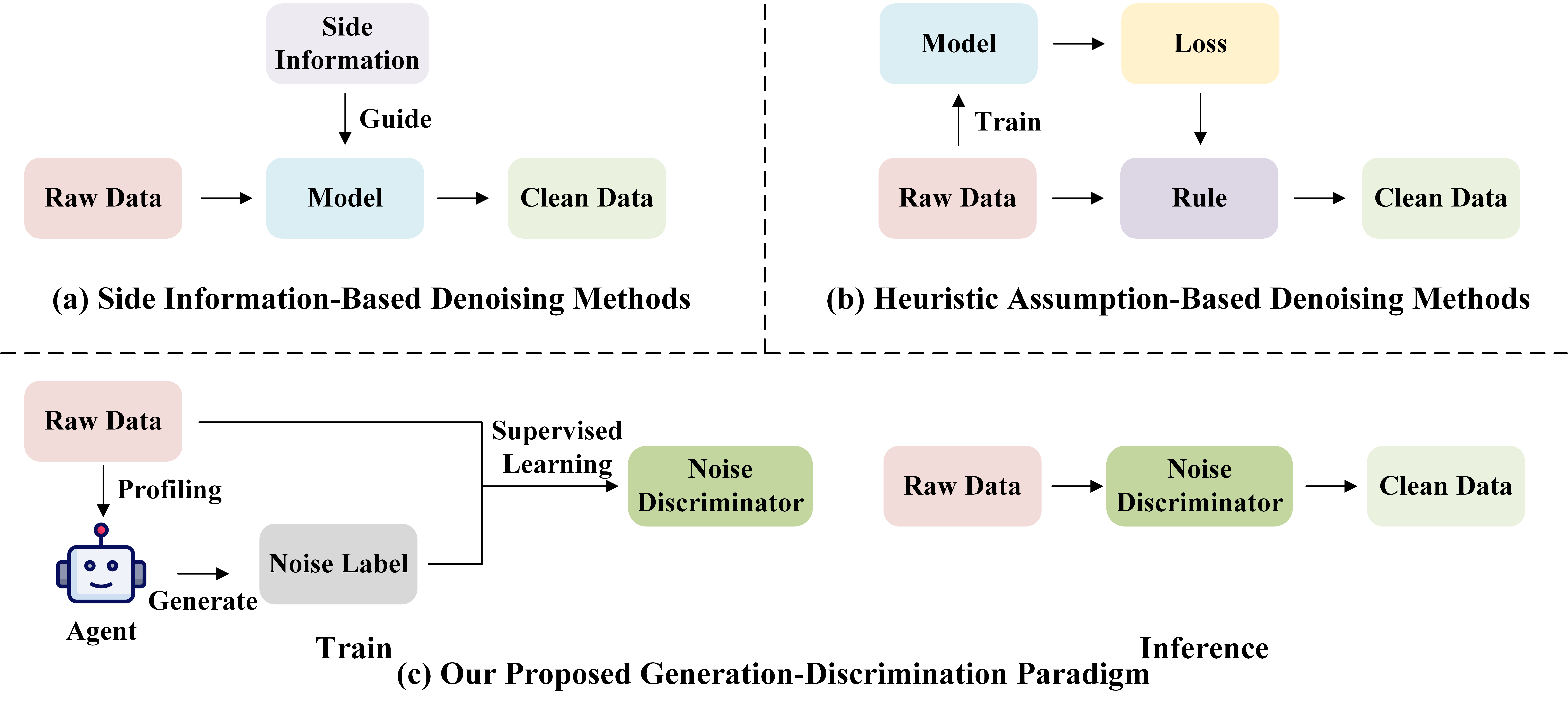}
\vspace{-2mm}
\caption{Comparison between traditional denoising methods and our proposed Generative-Discriminative paradigm.}
\label{fig:intro}
\vspace{-2mm}
\end{figure}

\section{Introduction}

Recommendation systems play a crucial role in mitigating information overload and providing personalized services, significantly enhancing user experience and boosting business revenue \cite{zhang2025dembr, zhang2025survey}. High-quality user feedback is fundamental to capturing user preferences and optimizing recommendation performance \cite{xin2023improving, wang2021denoising}. While explicit feedback (e.g., ratings) offers high reliability, its scarcity necessitates the widespread adoption of implicit feedback (e.g., clicks and purchases) due to its accessibility \cite{wang2021denoising, gao2022self, sun2021does}. However, the utility of implicit feedback is often compromised by inherent noise \cite{wang2021denoising}, as observed interactions do not always align with genuine user interests. For instance, users may click on items that do not align with their preferences due to herd behavior, curiosity, or accidental clicks. This noise undermines the accuracy of implicit feedback in reflecting users' true preferences, emphasizing the need for effective noise identification and filtering \cite{sun2021does}.

Despite the urgency of recommendation denoising, this task is severely hindered by the lack of explicit noise annotations in real-world implicit feedback. Given the unavailability of ground truth labels distinguishing valid interactions from noise, recommendation denoising inherently operates within an unsupervised setting. Compelled by this lack of supervision, existing methods typically resort to surrogate signals to infer noise patterns, generally falling into two categories: (1) \textbf{Side Information-based denoising methods}, which utilize external data to assist in noise identification, as illustrated in Figure \ref{fig:intro} (a). Early methods focused on dwell time and attention patterns \cite{buscher2009segment, fu2010towards, zhao2016gaze}, while later approaches introduced auxiliary behaviors \cite{han2024efficient, xin2023improving}, social networks \cite{sun2025model, quan2023robust}, and knowledge graphs \cite{tang2024editkg, zhu2023knowledge}. However, the data collection cost for such methods is high and may pose privacy risks \cite{wang2021denoising, wang2025ruleagent}. Additionally, external signals may introduce irrelevant information, potentially leading to further noise \cite{gao2022self, zhu2023knowledge}. (2) \textbf{Heuristic assumption-based denoising methods}, which rely on expert-defined empirical assumptions as denoising rules, as shown in Figure \ref{fig:intro} (b). For example, samples with high training loss \cite{wang2021denoising} or significant loss fluctuation across different models \cite{wang2022learning} are considered noise. However, these methods rely heavily on specific assumptions that require manual setup and trial-and-error, leading to high time and labor costs \cite{wang2025unleashing}. Moreover, these assumptions are not always valid and are difficult to generalize to other datasets, severely limiting the scalability and practical applicability of such methods \cite{wang2025ruleagent}. 

To overcome the inherent limitations of existing methods, we propose a paradigm shift grounded in a key insight: 'The best way to predict noise is to create it.' Instead of passively inferring noise from unlabeled data, we aim to proactively synthesize noise labels to reformulate denoising from an unsupervised filtering problem into a Creation-Recognition supervised learning problem. Achieving this goal necessitates a modeling approach capable of realistically simulating complex user behaviors. Fortunately, recent research on 'LLM-as-User' has demonstrated that agents powered by Large Language Models (LLMs) possess exceptional capabilities in semantic reasoning and human-like behavior simulation \cite{zhang2024generative, zhang2025survey}. Inspired by this potential, we propose to leverage LLM-empowered agents to simulate diverse non-preference-driven interactions. By accurately simulating the underlying mechanisms of how noise arises in real user interactions, we can construct a rich dataset of supervised noise labels that approximates real-world distributions. These explicit labels subsequently empower the training of a specialized noise recognizer, enabling effective identification of noise patterns without relying on costly side information or rigid heuristic assumptions.

Building on this reformulation, we propose the Creation-Recognition paradigm, a supervised denoising framework that actively creates noise through simulated user-system interactions and then recognizes noise through a trained parametric model, as illustrated in Figure \ref{fig:intro} (c). This paradigm differs fundamentally from conventional denoising methods: rather than designing increasingly complex rules to infer unknown noise after interactions have been observed, it first constructs labeled noisy interactions and then learns a recognizer from these labels. To instantiate this paradigm, we introduce \textbf{A}gentic \textbf{N}oise \textbf{C}reation framework for \textbf{H}uman simulation and den\textbf{O}ising \textbf{R}ecommendation (ANCHOR). By synthesizing high-quality noise labels through agent-based user simulation, ANCHOR empowers a dedicated noise recognizer to precisely identify noisy interactions within implicit feedback data, thereby significantly enhancing the robustness and performance of downstream recommendation models. ANCHOR implements this paradigm through a dual-stage process involving noise creation and noise recognition. Specifically, during the noise creation phase, ANCHOR simulates realistic user-system interactions by introducing a recommender-in-the-loop agentic framework. This framework synthesizes two distinct categories of noise: diverse out-of-preference noise and informative boundary-adjacent noise. For out-of-preference noise, we design five specialized simulation mechanisms (misclick, curiosity-driven noise, caption-biased noise, popularity-biased noise, and position-biased noise) to approximate real-world noise distributions. These mechanisms capture typical non-preference-driven interactions that should be filtered out. Furthermore, recognizing that not all noise is equally obvious, we also focus on boundary-adjacent noise, which refers to preference-adjacent interactions that share superficial similarities with user interests but do not reflect stable long-term intent. To synthesize such subtle noise, we introduce a creator-recognizer iterative refinement mechanism. In this process, the noise creator generates ambiguous interactions to challenge the recognizer, while the recognizer provides feedback on which cases are easily detected and which remain difficult. Through prompt-level reflection and memory updates, the creator gradually refines its generation strategy to produce more informative hard cases, thereby enriching the supervision available for noise recognition. In the noise recognition stage, we capitalize on the rich, labeled dataset constructed by ANCHOR to train a recommendation-oriented, semantics-aware recognizer capable of identifying noise patterns. This direct supervision enables the model to learn precise decision boundaries, effectively distinguishing complex noise patterns from genuine user interests. Using the trained noise recognizer, we identify and remove noise from the dataset and use the purified interactions to train downstream recommendation models, thereby improving recommendation performance. Meanwhile, the detected boundary-adjacent noise serves as boundary-sensitive supervisory signals, guiding the recommendation model to better capture user preference boundaries. In summary, our work contributes a paradigm-level reformulation of recommendation denoising together with three concrete technical advances and extensive empirical validation:
\vspace{-\topsep}

\begin{itemize} [leftmargin=*]
    \item \textbf{Paradigm-level reformulation.} We propose a Creation-Recognition supervised denoising paradigm, instantiated by ANCHOR, which proactively creates labeled noisy interactions and trains a dedicated recognizer for noise identification. This paradigm reformulates recommendation denoising from heuristic or unsupervised filtering into a supervised learning problem.

    \item \textbf{Structured out-of-preference noise simulation.} We design a structured noise creation framework based on LLM-as-User and recommender-in-the-loop User-Recommender Interaction Simulation. ANCHOR simulates five behavior-grounded out-of-preference noise types, including misclick, curiosity-driven, caption-biased, popularity-biased, and position-biased interactions.

    \item \textbf{Boundary-adjacent hard-noise construction.} We introduce a creator-recognizer iterative refinement mechanism to synthesize boundary-adjacent noise. By allowing the creator to generate preference-adjacent hard cases and refine its strategy using recognizer feedback, ANCHOR enriches the supervision signals with more informative and difficult noisy interactions.

    \item \textbf{Reusable semantic-collaborative recognizer.} We develop a reusable parametric noise recognizer that integrates collaborative ID signals and semantic representations. The generated noisy interactions serve as supervision for learning a transferable denoising module that can be applied to real interaction data and generalized beyond the simulated users.

    \item \textbf{Extensive empirical validation.} We conduct comprehensive experiments on three benchmark datasets and two recommendation backbones, including comparisons with state-of-the-art denoising baselines, ablation studies, hyperparameter sensitivity analysis, and robustness evaluation under different noise ratios. The results demonstrate that ANCHOR consistently improves recommendation robustness and performance.
\end{itemize}

\section{Related Work}
\subsection{Denoising in Recommendation}
In implicit‐feedback-based recommendation systems, observed interactions are typically interpreted as positive signals, whereas unobserved ones are treated as negatives \cite{ding2019sampler, gao2022self}. However, recent studies indicate that implicit feedback can be influenced by external factors, such as users' curiosity and herd behavior. These influences may introduce substantial noise and consequently degrade the performance of recommendation models \cite{wang2021denoising, sun2021does}. Existing denoising approaches can be broadly grouped into the following category: \textbf{1) Side Information–Based Methods}, which utilize external data as auxiliary signals to identify noisy interactions. Early studies \cite{buscher2009segment, fu2010towards, zhao2016gaze} relied on dwell time and attention patterns, whereas later work incorporated auxiliary behaviors \cite{han2024efficient, xin2023improving}, social networks \cite{sun2025model, quan2023robust}, and knowledge graphs \cite{tang2024editkg, zhu2023knowledge} to enhance noise detection. Although effective, these methods often entail substantial data acquisition costs and raise privacy concerns \cite{wang2021denoising, wang2025ruleagent}. Moreover, the introduced external signals may contain task-irrelevant information, potentially contributing additional noise \cite{gao2022self, zhu2023knowledge}. \textbf{2) Heuristic Assumption–Based Methods} rely on expert-designed empirical rules. For example, T-CE \cite{wang2021denoising} assumes that samples with high training loss are likely to be noisy, while DeCA \cite{wang2022learning} is based on the assumption that samples exhibiting large loss fluctuations across multiple models tend to be unreliable. Other methods follow similar principles. SGDL \cite{gao2022self} introduces the prior that clean samples emerge early in training and therefore assigns adaptive weights according to sample similarity. BOD \cite{wang2023efficient} further assumes that samples yielding consistent losses under different loss functions are more likely to be clean. However, these approaches depend heavily on manually crafted heuristics and extensive trial-and-error, which are both time-consuming and labor-intensive \cite{wang2025unleashing}. In addition, their underlying assumptions do not always hold and often fail to generalize across datasets, significantly limiting their scalability and practical applicability \cite{wang2025ruleagent}.

Fundamentally, the limitations of existing denoising methods arise from the lack of reliable noise annotations in implicit feedback, which prevents models from accurately learning the underlying noise distribution from real user interactions. In contrast, our approach explicitly simulates the noise-generation mechanisms in implicit feedback through agent-based user behavior modeling. This enables the construction of reliably labeled noisy interactions and allows the denoising model to learn the true noise distribution in a fully data-driven manner.

\subsection{LLM-powered Agents in Recommendation}
With the rise of LLM-powered agents equipped with advanced reasoning, planning, and decision-making capabilities \cite{argyle2023out, rana2023sayplan}, recent studies have begun to explore their integration into recommendation systems \cite{zhang2025survey}. Existing work can be broadly classified into two directions: \textbf{1) Agent as Recommender.} This line of research treats the agent itself as the recommendation system, leveraging its strong semantic understanding to model user interaction histories and incorporate item semantics for generating personalized recommendations. For instance, RecMind \cite{wang2024recmind} constructs an LLM-based recommender agent that performs personalized recommendation through structured planning and tool use, achieving strong performance across diverse tasks. MACRec \cite{wang2024macrec} further proposes a collaborative multi-agent LLM framework in which specialized agents jointly address recommendation problems. \textbf{2) Agent as User Simulator.} This line of work leverages agents’ ability to model human behavior to simulate user actions in recommendation scenarios, such as expressing preferences or providing feedback. For example, Agent4Rec \cite{zhang2024generative} constructs a user agent with profiling, memory, and action modules to emulate behaviors including viewing, rating, and feedback generation.

Despite their promising capabilities, both lines of agent-based research largely overlook the inherent noise in implicit feedback, which can distort user preference modeling and ultimately degrade recommendation accuracy. In contrast, our work diverges from these paradigms by employing agents not as recommenders, but as noise generators that mimic non–preference-driven behaviors in implicit feedback. By generating labeled noisy interactions and using them to supervise a noise recognizer, our framework effectively identifies and filters out noise from the training data, resulting in more reliable user preference signals and improved recommendation performance.

\section{Problem Formulation}
\textbf{Recommendation Denoising.} We study recommendation denoising under the standard implicit-feedback setting. Consider the user--item interaction set $\mathcal{D} = \{(u, i, r_{u,i}) \mid u \in \mathcal{U},\, i \in \mathcal{I}\}$, where $r_{u,i} \in \{0,1\}$ denotes whether user $u$ has an observed interaction with item $i$. Following common implicit-feedback recommendation protocols, we treat observed user--item records as positive implicit interactions and do not use explicit rating values as preference labels. Here, $\mathcal{U} \in \mathbb{R}^{|\mathcal{U}|}$ is the user set, and $\mathcal{I} \in \mathbb{R}^{|\mathcal{I}|}$ is the item set. Recommendation models typically utilize the interaction data $\mathcal{D}$ to learn latent representations for users and items, namely $\mathbf{Z}_{\mathcal{U}} \in \mathbb{R}^{|\mathcal{U}| \times d}$ and $\mathbf{Z}_{\mathcal{I}} \in \mathbb{R}^{|\mathcal{I}| \times d}$, where $d$ is the representation dimension. A recommendation model $f$ parameterized by $\theta$ is then trained to estimate interaction likelihoods. Formally, the training objective is given by:
\begin{equation}
\theta^{*} = \arg\min_{\theta} \mathcal{L}_{rec}(\mathcal{D}),
\label{optimization objective}
\end{equation}
where $\theta^{*}$ represents the optimal parameters of $f$, and $\mathcal{L}_{rec}$ denotes the recommendation loss, instantiated here with the BPR objective \cite{rendle2012bpr}:
\begin{equation}
\mathcal{L}_{rec} =
\mathbb{E}_{(u,i,j)\in\mathcal{D}}
\left[-\log\big(\sigma(f(z_u, z_i) - f(z_u, z_j))\big)\right],
\label{bprloss}
\end{equation}
where $(u,i,j)$ includes a user $u$, an interacted positive item $i$, and a non-interacted negative item $j$. The function $\sigma(\cdot)$ denotes the sigmoid activation. While $r_{u,i} = 1$ typically suggests that user $u$ prefers item $i$, real-world implicit interactions (such as views, purchases) may be influenced by curiosity, herd behavior, accidental clicks, and other non–preference-driven factors. These effects introduce substantial noise, causing observed interactions to deviate from users’ true underlying preferences. The objective of the recommendation denoising task is to identify the noisy interactions within the observed interaction matrix $\mathcal{D}$, denoted by the noise set $\mathcal{N}$, and remove them to obtain a clean interaction matrix $\mathcal{D}^{*} = \mathcal{D} \setminus \mathcal{N} \in \{0,1\}^{|\mathcal{U}| \times |\mathcal{I}|}$, which better reflects users’ genuine preference signals.

\noindent\textbf{Our Proposed Creation-Recognition Paradigm.} Since the elements of $\mathcal{N}$ are not explicitly labeled, directly learning to distinguish noisy interactions from true preference signals is inherently challenging. To address this, our creation-recognition paradigm simulates non-preference-driven behaviors in implicit feedback (e.g., misclicks, curiosity-driven views, herd behavior). This process yields a synthetic noise set $\widetilde{\mathcal{N}} = \{(u,i,y_{u,i}) \mid y_{u,i}=1\}$, whose distribution is designed to approximate the latent noise set $\mathcal{N}$. The availability of $\widetilde{\mathcal{N}}$ enables supervised training of a noise recognizer $g$ with parameters $\phi$, which assigns each interaction a noise probability $s_{u,i} = g_\phi(u,i)$. Formally, the training objective of the recognizer is defined as:
\begin{equation}
\phi^{*} = 
\arg\min_{\phi} \ \mathcal{L}_{\text{noise}}(\phi;\widetilde{\mathcal{N}},\mathcal{D}),
\end{equation}
where $\phi^{*}$ represents the optimal parameters of $g$, $\mathcal{L}_{\text{noise}}$ denotes the optimization objective, formulating noise recognition as a binary classification problem. Using the learned recognizer, we estimate the noise set in the original interactions as $\widehat{\mathcal{N}} = \{(u,i)\mid g_{\phi^{*}}(u,i) > \tau\}$, with $\tau$ being a confidence threshold. The clean interaction matrix is then obtained by $\mathcal{D}^{*} = \mathcal{D} \setminus \widehat{\mathcal{N}}.$

\begin{figure*}[t]
\centering
\includegraphics[width=0.96\linewidth]{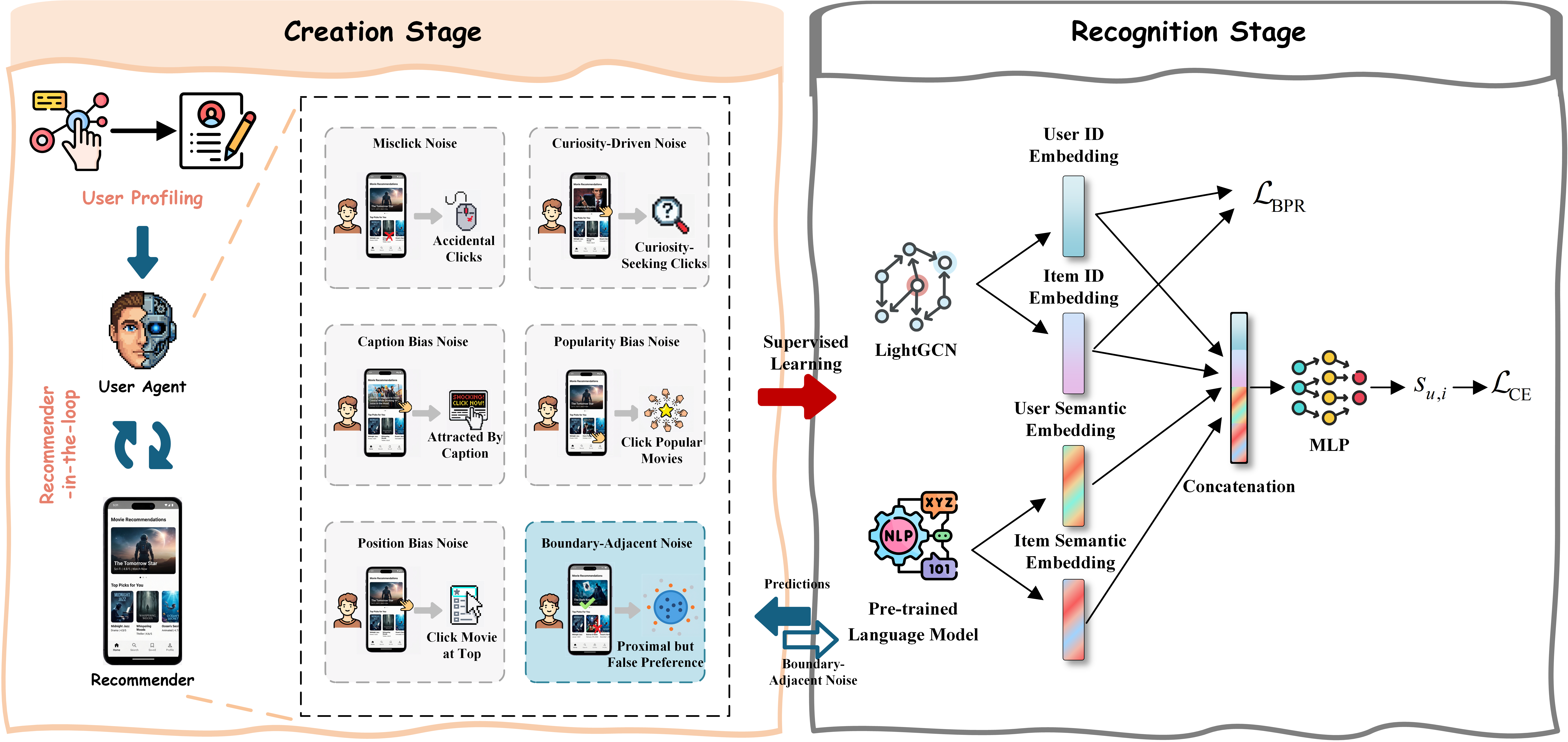}
\vspace{-3mm}
\caption{The overall framework of the proposed ANCHOR framework.}
\label{fig:overall}
\vspace{-2mm}
\end{figure*}

\section{Methodology}
ANCHOR consists of three core modules that instantiate the Creation-Recognition paradigm: User Simulation-Driven Out-of-Preference Noise Generation, Creator-Recognizer Iterative Boundary Refinement, and Supervised Learning of a Reusable Noise Recognizer. Specifically, the first module introduces a recommender-in-the-loop agentic framework to simulate realistic user-system interactions and synthesize diverse out-of-preference noise through five behavior-grounded mechanisms. The second module focuses on boundary-adjacent hard noise through a feedback-driven interaction between the noise creator and the noise recognizer. Finally, the third module uses the synthesized labels to train a recommendation-oriented, semantics-aware parametric recognizer. This process enables precise identification of noise patterns, transforming recommendation denoising from an unsupervised challenge into a fine-grained supervised task that significantly enhances the robustness and performance of downstream recommendation models.
\subsection{User Simulation-Driven Out-of-Preference Noise Generation}
\subsubsection{Agent-based User Profiling and Simulation}

To realistically simulate non-preference-driven user behaviors, ANCHOR leverages both the semantic reasoning and human behavior simulation capabilities of LLM-based Agents \cite{wang2025user, hu2023unlocking} to profile individual users. By deeply interpreting the semantic characteristics within historical behaviors, our approach enables the agent to form a coherent representation of user preferences and behavioral tendencies. This representation serves as the basis for generating behaviorally plausible interactions under realistic recommendation scenarios.

Given a user $u \in \mathcal{U}$, we denote the set of items previously interacted with by $u$ as $\mathcal{I}_u = \{ i \mid r_{u,i}=1 \}$. Each item $i \in \mathcal{I}_u$ is associated with textual content $\mathbf{t}_i$, such as titles, descriptions, or metadata. These textual signals collectively reflect the semantic characteristics of the user's historical interactions. We concatenate the textual content of the interacted items and feed it into an LLM (e.g., GPT-3.5 Turbo) to derive a high-level behavioral profile of the user. Formally, the user profiling process is defined as:
\begin{equation}
\mathbf{p}_u = f_{\text{LLM}}\big(\{\mathbf{t}_i \mid i \in \mathcal{I}_u\}\big),
\label{eq:user_profile}
\end{equation}
where $f_{\text{LLM}}(\cdot)$ denotes the LLM-based user understanding function (the specific prompt is provided in Appendix \ref{app:user_profiling_prompts}), and $\mathbf{p}_u$ represents the inferred profile capturing user preferences, behavioral tendencies, and decision-making patterns. This inferred profile $\mathbf{p}_u$ effectively functions as a personalized cognitive blueprint for the user agent, capturing the nuanced behavioral tendencies necessary to drive realistic interactions. Consequently, this profound understanding empowers the agent to generate dynamic interaction sequences that naturally reproduce the complex noise patterns characteristic of real-world user feedback.
To authentically simulate user-system interactions, we establish a recommender-in-the-loop agentic framework where an independent recommendation system supplies candidate items to user agents, creating a realistic interaction environment. Specifically, for each user $u$, we utilize a recommendation model, instantiated as a LightGCN pre-trained on historical data, to produce a ranked list of candidate items $\mathcal{C}_u \subset \mathcal{I}$. The user agent observes $\mathcal{C}_u$ and determines whether to simulate a noisy interaction with an item based on both the inferred profile $\mathbf{p}_u$ and simulated non-preference-driven behavioral mechanisms.

\subsubsection{Construction of Out-of-Preference Noise}
Based on the user profiles and the candidate item exposure set, ANCHOR constructs diverse out-of-preference noise by explicitly simulating non-preference-driven user behaviors. These behaviors correspond to common deviations observed in real-world implicit feedback, where interactions are triggered by factors other than users’ genuine long-term interests. Formally, given a user $u$ with profile $\mathbf{p}_u$ and candidate item set $\mathcal{C}_u$, we generate synthetic noisy interactions by prompting the user agent to select items from $\mathcal{C}_u$ under specific behavioral instructions. All generated noisy items are required to satisfy $i \in \mathcal{C}_u$ and $i \notin \mathcal{I}_u$, ensuring that noise interactions are exposure-consistent and do not overlap with historical interactions. We explicitly model five distinct behavioral mechanisms that lead to out-of-preference noise:

\noindent \textbf{Misclick Noise.}
Misclick noise models accidental interactions caused by interface confusion, inattentive operations, or unintended clicks. Such interactions are assumed to be independent of user preferences \cite{ren2023disentangled}. To simulate this behavior, the user agent is instructed to randomly select items from the candidate set $\mathcal{C}_u$ without considering semantic relevance to $\mathbf{p}_u$. The resulting interactions are diverse and reflect the stochastic nature of accidental clicks. These interactions constitute the most preference-irrelevant form of noise. 


\noindent \textbf{Curiosity-Driven Noise.}
Curiosity-driven noise captures exploratory behaviors in which users intentionally interact with items outside their core interests in order to discover new content. Unlike misclicks, these interactions are weakly related to user preferences but do not reflect stable preference signals. To generate this noise, the agent leverages the user profile $\mathbf{p}_u$ to identify items in $\mathcal{C}_u$ that are semantically adjacent yet distinct from the user’s dominant interest patterns. This process simulates rational but exploratory clicks driven by novelty-seeking rather than preference satisfaction. 


\noindent \textbf{Caption Bias Noise.}
Caption bias noise captures interactions driven by superficial textual attributes, such as catchy titles, trending entities, or intriguing descriptions, regardless of the item’s actual relevance \cite{hofmann2012caption}. To model this phenomenon, the agent examines the textual content of candidate items and selects those with high surface-level appeal, even if they are weakly related or unrelated to $\mathbf{p}_u$. These interactions reflect impulsive decisions influenced by presentation bias rather than intrinsic preference. 


\noindent \textbf{Popularity Bias Noise.}
Popularity bias noise represents interactions driven by social proof, hype, or global popularity. In such scenarios, users are inclined to interact with items simply because they are widely consumed or highly recommended by others \cite{klimashevskaia2024survey}. To simulate this behavior, we augment the agent’s input with item popularity statistics derived from the dataset. The agent is guided to preferentially select highly popular items from $\mathcal{C}_u$, regardless of their alignment with the user’s personal preferences. This models herd behavior commonly observed in large-scale recommendation systems. 

\noindent \textbf{Position Bias Noise.}
Position bias noise models the tendency of users to click items appearing at prominent positions in a ranked recommendation list \cite{craswell2008experimental}. Such interactions are driven by exposure order rather than preference relevance. To generate this noise, the agent is provided with the ranked candidate list along with item position indices. Items appearing at higher ranks are assigned higher click likelihoods, enabling the agent to generate interactions primarily motivated by ranking position instead of semantic match with $\mathbf{p}_u$.

For detailed implementation, the specific prompts guiding the simulation of these five noise types are comprehensively listed in Appendix \ref{app:noise_creation_prompts}. The first three noise types, namely misclick, curiosity-driven, and caption bias, share a unified generation mechanism: the user agent selects items from the candidate set based on the user profile and a behavior-specific instruction, without relying on external signals beyond item semantics. Formally, for a noise type $k \in \{\text{Misclick}, \text{Curiosity}, \text{Caption}\}$, the synthetic noisy interactions are generated as:
\begin{equation}
\widetilde{\mathcal{N}}^{(k)}_u
=
f_{\text{LLM}}\!\left(
\mathbf{p}_u,\;
\mathcal{C}_u,\;
\boldsymbol{\psi}^{(k)}
\right),
\label{eq:basic_noise_generation}
\end{equation}
where $\boldsymbol{\psi}^{(k)}$ denotes the prompt instruction encoding the corresponding non-preference-driven behavioral constraint. The output $\widetilde{\mathcal{N}}^{(k)}_u$ is a set of item interactions satisfying $i \in \mathcal{C}_u$ and $i \notin \mathcal{I}_u$, representing synthetic out-of-preference noise of type $k$ for user $u$.

To account for popularity-driven behavior, we further condition the agent on item-level popularity signals. Let $\mathbf{q}_i$ denote the global popularity statistic of item $i$, such as the total number of interactions in the training dataset. The popularity-biased noise is generated as:
\begin{equation}
\widetilde{\mathcal{N}}^{(\text{Popularity})}_u
=
f_{\text{LLM}}\!\left(
\mathbf{p}_u,\;
\mathcal{C}_u,\;
\{\mathbf{q}_i \mid i \in \mathcal{C}_u\},\;
\boldsymbol{\psi}^{(\text{Popularity})}
\right),
\label{eq:popularity_noise_generation}
\end{equation}
where the popularity statistics guide the agent to preferentially select globally popular items, simulating interactions driven by social proof rather than personalized relevance.
Similarly, position bias is modeled by conditioning the agent on the ranked exposure positions of candidate items. Let $\text{pos}_u(i)$ denote the position index of item $i$ in the recommendation list shown to user $u$. The position-biased noise is generated as:
\begin{equation}
\widetilde{\mathcal{N}}^{(\text{Position})}_u
=
f_{\text{LLM}}\!\left(
\mathbf{p}_u,\;
\mathcal{C}_u,\;
\{\text{pos}_u(i) \mid i \in \mathcal{C}_u\},\;
\boldsymbol{\psi}^{(\text{Position})}
\right),
\label{eq:position_noise_generation}
\end{equation}
where items with higher exposure positions are assigned higher selection likelihoods, capturing clicks induced by ranking prominence rather than preference alignment.

Collectively, the five noise generation processes produce a unified synthetic out-of-preference noise set. Formally, for each user $u$, the out-of-preference noise set is defined as the union of the five behavior-specific noise sets:
\begin{equation}
\begin{aligned}
\widetilde{\mathcal{N}}_u
&=
\bigcup_{k \in \mathcal{K}}
\widetilde{\mathcal{N}}^{(k)}_u, \\
\mathcal{K}
=
\{\text{Misclick},\ \text{Curiosity}&,\ \text{Caption},\ \text{Popularity},\ \text{Position}\},
\end{aligned}
\label{eq:out_of_preference_union}
\end{equation}
and the overall synthetic out-of-preference noise set is obtained by aggregating over a sampled subset of users. Specifically, to ensure computational efficiency and scalability, we sample a user subset $\widetilde{\mathcal{U}} \subset \mathcal{U}$ (e.g., $|\widetilde{\mathcal{U}}| = 1,000$) and construct:

\begin{equation}
\widetilde{\mathcal{N}}
=
\bigcup_{u \in \widetilde{\mathcal{U}}}
\widetilde{\mathcal{N}}_u,
\qquad
\widetilde{\mathcal{U}} \subset \mathcal{U}.
\label{eq:global_out_of_preference_noise}
\end{equation}
This synthesized noise set approximates the distribution of non-preference-driven interactions in real-world implicit feedback and provides reliable supervision for learning the noise recognizer in subsequent stages.

\subsection{Creator-Recognizer Iterative Boundary Refinement}
In this work, we further consider a challenging category of noisy interactions termed boundary-adjacent noise. These interactions are preference-adjacent in the sense that they may share superficial semantic similarities with a user's historical interests, but they still fail to reflect the user's stable long-term intent. Therefore, they serve as hard noisy cases that are more difficult to distinguish from genuine preference-driven interactions than clearly out-of-preference noise. Identifying such cases can provide fine-grained supervision for improving the recognizer's sensitivity to ambiguous preference-noise patterns. However, these interactions are difficult to capture using static heuristics because their misleading surface relevance makes them resemble positive signals.


To address this challenge, we propose a creator-recognizer iterative boundary refinement mechanism. The core idea is to construct boundary-adjacent noise through feedback-driven interaction between two components. Importantly, this process is not a strict gradient-based minimax optimization or a classical GAN-style training procedure. Instead, the creator is refined through prompting, reflection, and memory updates based on recognizer feedback. Concretely, the refinement process consists of four steps in each iteration. First, the creator generates boundary-adjacent candidate interactions conditioned on the user history, user profile, candidate items, and the current refinement memory. Second, the recognizer scores these candidates and estimates their noise probabilities. Third, the scoring results are converted into structured feedback, where low-score noisy samples are treated as hard cases and high-score noisy samples are treated as easy cases. Finally, the creator reflects on this feedback and updates its refinement memory, which is then used to guide the next generation round. This loop allows the generated samples to become progressively more informative without updating the parameters of the LLM creator.
\begin{itemize}[leftmargin=*]
  \item A \textbf{noise creator} that synthesizes preference-adjacent noisy interactions based on user profiles, candidate items, and accumulated feedback from previous rounds.
  \item A \textbf{noise recognizer} (detailed in Section \ref{sec:recognizer}) that evaluates these generated cases and provides feedback indicating which samples are easily detected and which remain difficult to distinguish from genuine preference-driven interactions.
\end{itemize}


At iteration $t$, the noise creator generates a set of boundary-adjacent candidate interactions, denoted by $\mathcal{N}^{(t)}$, conditioned on the user history, preference summaries, candidate items, and the accumulated refinement memory. The prompts of this process can be found in Appendix \ref{app:adversarial_prompts}. The generated interactions are then evaluated by the noise recognizer $g_\phi$, which assigns each interaction $(u,i)$ a noise probability $s_{u,i}$ as defined in Eq.~\eqref{eq:noise_prediction}. Based on the recognizer's predictions, especially the cases that are assigned low noise probabilities and are therefore difficult to detect, the creator receives structured feedback for reflection and strategy refinement. The recognizer is updated using the supervised noise learning objective in Eq.~\eqref{eq:disc_total_loss}, while progressively incorporating the newly generated boundary-adjacent noise into its training set. Through this iterative interaction, the recognizer is exposed to increasingly informative hard cases, improving its sensitivity to ambiguous patterns between genuine preference and non-preference-driven interactions.



Rather than optimizing a strict minimax objective, we implement the refinement process as an iterative feedback loop. Let $t$ denote the index of the current refinement iteration, and let $\mathcal{M}^{(t)}$ denote the refinement memory at iteration $t$, which stores the recognizer feedback and the summarized hard/easy patterns from previous rounds. We denote the noise creator by $G$. Given the interaction history $\mathcal{H}_u$ of user $u$, the user profile $\mathbf{p}_u$, the candidate item set $\mathcal{C}_u$, and the current refinement memory $\mathcal{M}^{(t)}$, the creator generates a set of boundary-adjacent noisy interactions $\mathcal{N}^{(t)}$ as:
\begin{equation}
\mathcal{N}^{(t)}
=
G\left(
\mathcal{H}_u,\,
\mathbf{p}_u,\,
\mathcal{C}_u,\,
\mathcal{M}^{(t)}
\right),
\label{eq:boundary_generation}
\end{equation}
where $\mathcal{H}_u$ denotes the user interaction history, $\mathbf{p}_u$ is the user profile, and $\mathcal{C}_u$ is the candidate item set. The generated set $\mathcal{N}^{(t)}$ consists of preference-adjacent noisy interactions produced in the $t$-th refinement iteration.

The recognizer then evaluates each generated interaction by assigning a noise probability:
\begin{equation}
s_{u,i}^{(t)} = g_\phi(u,i), \quad (u,i) \in \mathcal{N}^{(t)},
\label{eq:boundary_scoring}
\end{equation}
where $g_\phi$ denotes the noise recognizer parameterized by $\phi$, and $s_{u,i}^{(t)}$ denotes the predicted noise probability for interaction $(u,i)$ at iteration $t$. Based on these scores, we construct recognizer feedback $\mathcal{F}^{(t)}$, which summarizes hard cases that receive low noise probabilities and easy cases that are confidently detected. The refinement memory is then updated through prompt-level reflection:
\begin{equation}
\mathcal{M}^{(t+1)}
=
\mathrm{Reflect}\left(
\mathcal{M}^{(t)},\,
\mathcal{F}^{(t)}
\right),
\label{eq:memory_update}
\end{equation}

The recognizer is then trained with the accumulated generated noise:
\begin{equation}
\phi^{(t+1)}
=
\arg\min_{\phi}
\mathcal{L}_{\text{noise}}
\left(
\phi;\,
\mathcal{D}_{\text{disc}} \cup \bigcup_{\ell=1}^{t}\mathcal{N}^{(\ell)}
\right),
\label{eq:recognizer_refinement}
\end{equation}
where $\ell$ indexes previous refinement iterations, $\bigcup_{\ell=1}^{t}\mathcal{N}^{(\ell)}$ denotes all boundary-adjacent noisy interactions generated up to iteration $t$, and $\mathcal{D}_{\text{disc}}$ is the recognizer training set constructed from original interactions and generated out-of-preference noise. This formulation emphasizes that the creator is refined through feedback, reflection, and memory updates, rather than through gradient-based optimization of LLM parameters.



In summary, the proposed creator-recognizer iterative refinement mechanism transforms boundary-adjacent noise from a disturbance into an informative supervisory signal. By repeatedly exposing the recognizer to preference-adjacent hard cases, this mechanism improves its ability to distinguish genuine preference-driven interactions from subtle non-preference-driven interactions. The identified boundary-adjacent noise can also be used by downstream recommendation models as hard negative signals, thereby facilitating more robust preference learning.
\subsection{Supervised Learning of Noise Recognizer}
\label{sec:recognizer}

Given the synthetic out-of-preference noise set $\widetilde{\mathcal{N}}$ and the boundary-adjacent noise generated through iterative refinement, we introduce how ANCHOR learns a supervised noise recognizer to model noise patterns in implicit feedback. The key objective of this stage is to transfer the agent-created supervision into a semantic-collaborative denoising module that can distinguish non-preference-driven interactions from genuine preference signals in real interaction data.
\paragraph{Training data construction.}
To avoid trivial shortcuts where the recognizer simply memorizes user identities as noisy users, we restrict recognizer training to the users involved in noise generation and include both their original interactions and generated noisy interactions. This construction forces the recognizer to learn interaction-level noise patterns from semantic and collaborative signals, rather than relying on whether a user appears in the generation stage. Formally, let $\widetilde{\mathcal{U}} \subset \mathcal{U}$ denote the sampled user subset used in agent-based noise simulation. We construct the recognizer training set by combining the original interactions of these users and the generated noise interactions:
\begin{equation}
\mathcal{D}_{\text{disc}}
=
\{(u,i,0)\mid (u,i)\in\mathcal{D},\, u\in\widetilde{\mathcal{U}}\}
\;\cup\;
\{(u,i,1)\mid (u,i,1)\in\widetilde{\mathcal{N}}\},
\label{eq:disc_training_set}
\end{equation}
where label $0$ indicates a potentially clean interaction and label $1$ denotes a synthetic noisy interaction. This construction forces the recognizer to rely on interaction-level signals rather than user-level identity cues when distinguishing noise.

\paragraph{Recognizer architecture.}
The noise recognizer $g_\phi$ is designed as a reusable parametric model that combines collaborative signals and semantic information. Specifically, for the recommendation-guided component, we employ a LightGCN model to learn ID-based latent representations for users and items, producing representations $z_u \in \mathbb{R}^d$ and $z_i \in \mathbb{R}^d$. These collaborative representations preserve the structural information in user--item interactions and stabilize recognizer learning through a standard BPR objective:
\begin{equation}
\mathcal{L}_{\text{BPR}}
=
\mathbb{E}_{(u,i,j)\in\mathcal{D}_{\text{disc}}}
\left[
-\log \sigma\big(f(z_u,z_i)-f(z_u,z_j)\big)
\right].
\label{eq:disc_bpr_loss}
\end{equation}

In parallel, to explicitly model noise characteristics, we introduce a noise classification head that integrates both ID-based and semantic representations. For each interaction $(u,i)$, we obtain semantic embeddings for the user and item by encoding the user profile $\mathbf{p}_u$ and item textual content $\mathbf{t}_i$ using a pre-trained language model \cite{wang2020minilm}, yielding $\mathbf{s}_u$ and $\mathbf{s}_i$, respectively. The recognizer then predicts the noise probability as:
\begin{equation}
s_{u,i}
=
g_\phi(u,i)
=
\mathrm{MLP}\!\left(
\left[z_u \,\|\, z_i \,\|\, \mathbf{s}_u \,\|\, \mathbf{s}_i\right]
\right),
\label{eq:noise_prediction}
\end{equation}
where $[\cdot\|\cdot]$ denotes vector concatenation. The noise classification head is trained using a binary cross-entropy loss \cite{ruby2020binary}:
\begin{equation}
\mathcal{L}_{\text{CE}}
=
\mathbb{E}_{(u,i,y_{u,i})\in\mathcal{D}_{\text{disc}}}
\left[
- y_{u,i}\log s_{u,i}
- (1-y_{u,i})\log(1-s_{u,i})
\right].
\label{eq:noise_ce_loss}
\end{equation}

\paragraph{Overall training objective.}
The final training objective of the noise recognizer jointly optimizes the noise classification loss and the BPR loss:
\begin{equation}
\mathcal{L}_{\text{noise}}
=
\mathcal{L}_{\text{CE}}
+
\alpha\,\mathcal{L}_{\text{BPR}},
\label{eq:disc_total_loss}
\end{equation}
where $\alpha$ controls the relative importance of collaborative signal preservation. By combining supervised noise recognition with recommendation-aware representation learning, the recognizer learns to identify non-preference-driven interactions while maintaining sensitivity to genuine user preferences. After training, $g_{\phi^*}$ serves as a reusable denoising module that can be applied to the original interaction data to estimate noisy interactions, rather than relying on repeated LLM prompting during deployment.

\section{Experiments}
We conduct comprehensive experiments to evaluate ANCHOR from four perspectives, corresponding to the following research questions:
\begin{itemize} [leftmargin=*]
\item \textbf{RQ1}: How does ANCHOR compare with state-of-the-art denoising methods when integrated into different recommendation backbones across diverse datasets?
\item \textbf{RQ2}: How do the key components of ANCHOR contribute to its overall effectiveness in noise identification and recommendation performance?
\item \textbf{RQ3}: How sensitive is ANCHOR to its key hyperparameters?
\item \textbf{RQ4}: How effectively does ANCHOR identify and remove noisy interactions ?
\end{itemize}
\subsection{Experimental Settings}
\begin{table}[htbp]
\caption{Statistics of the datasets.}
  \centering
    \begin{tabular}{l|r|r|r}
    \hline
    \textbf{Dataset} & \multicolumn{1}{|c}{\textbf{DBbook2014}} & \multicolumn{1}{|c}{\textbf{Book-Crossing}} & \multicolumn{1}{|c}{\textbf{MovieLens-1M}} \\
    \hline
    \#Users & 5,576 & 6,616 & 6,040 \\
    \#Items & 2,680 & 8,853 & 3,260 \\
    \#Interactions & 65,961 & 110,662 & 998,539 \\  
    \hline
    \end{tabular}%
  \label{tab:statistics_datasets}%
\end{table}%

\subsubsection{\textbf{Dataset Description}}
We conduct extensive experiments on three real-world datasets: DBbook2014 \cite{cao2019unifying}, Book-Crossing \cite{dong2017hybrid}, and MovieLens-1M \cite{noia2016sprank}. Crucially, all three datasets provide essential textual information regarding items, which is indispensable for enabling the semantic reasoning capabilities of our agentic framework. Following the standard implicit-feedback recommendation setting, we convert observed user--item records into binary interactions and use them to model whether an interaction has occurred. Following previous work \cite{hu2025bridging}, we preprocess the data by filtering out users with fewer than 10 interactions in MovieLens-1M, and fewer than 5 interactions in both DBbook2014 and Book-Crossing. The descriptive statistics after implicit-feedback preprocessing are summarized in Table~\ref{tab:statistics_datasets}.
\subsubsection{\textbf{Baselines}}
The primary objective of this work is to enhance the performance of existing recommendation models by mitigating noise in user feedback. To this end, we select two representative implicit feedback models to serve as backbones: GMF \cite{he2017neural} and LightGCN \cite{he2020lightgcn}. We compare ANCHOR against five representative recommendation denoising methods: T-CE \cite{wang2021denoising}, DeCA \cite{wang2022learning}, BOD \cite{wang2023efficient}, DCF \cite{he2024double}, and LLaRD \cite{wang2025unleashing}.

\begin{table*}[t]
\centering
\caption{Overall performance comparison of different baselines on the backbone models. Bold numbers indicate the best performance, and underlined numbers indicate the second-best performance. "R" and "N" stand for Recall and NDCG, respectively.}
\vspace{-2mm}
\label{table:main_all}
\resizebox{0.98\textwidth}{!}{%
\begin{tabular}{clcccccccccccc}
\toprule
\multicolumn{2}{c}{\textbf{Dataset}} & \multicolumn{4}{c}{\textbf{DBbook2014}} & \multicolumn{4}{c}{\textbf{Book-Crossing}} & \multicolumn{4}{c}{\textbf{MovieLens-1M}} \\ 
\cmidrule(r){1-2} \cmidrule(r){3-6} \cmidrule(r){7-10} \cmidrule(r){11-14}
\textbf{Backbone} & \textbf{Method} & \textbf{R@10} & \textbf{N@10} & \textbf{R@20} & \textbf{N@20} & \textbf{R@10} & \textbf{N@10} & \textbf{R@20} & \textbf{N@20} & \textbf{R@10} & \textbf{N@10} & \textbf{R@20} & \textbf{N@20} \\ 
\midrule
\midrule
\multirow{7}{*}{\makecell{GMF}} 
& Normal & 0.1565 & 0.1170 & 0.2271 & 0.1410 & 0.0378 & 0.0285 & 0.0562 & 0.0343 & 0.1386 & 0.3430 & 0.2249 & 0.3512 \\
& T-CE    & 0.1642 & 0.1205 & 0.2356 & 0.1498 & 0.0401 & 0.0312 & 0.0605 & 0.0388 & 0.1465 & 0.3502 & 0.2318 & 0.3555 \\
& DeCA    & 0.1823 & 0.1384 & 0.2619 & 0.1682 & 0.0520 & 0.0415 & 0.0764 & 0.0502 & 0.1408 & 0.3555 & 0.2308 & 0.3521 \\
& BOD     & 0.1859 & 0.1442 & 0.2541 & 0.1682 & 0.0592 & 0.0483 & 0.0841 & 0.0560 & 0.1498 & 0.3611 & 0.2314 & 0.3763 \\
& DCF    & 0.1905 & 0.1477 & 0.2495 & 0.1751 & 0.0568 & 0.0507 & 0.0833 & 0.0568 & 0.1486 & 0.3610 & 0.2293 & 0.3732 \\
& LLaRD  & \underline{0.1972} & \underline{0.1533} & \underline{0.2618} & \underline{0.1799}  & \underline{0.0694} & \underline{0.0556} & \underline{0.0931} & \underline{0.0668} & \underline{0.1755} & \underline{0.3693} & \underline{0.2615} & \underline{0.3824} \\
\cmidrule{2-14}
& \textbf{ANCHOR} & \textbf{0.2155} & \textbf{0.1712} & \textbf{0.3007} & \textbf{0.2043} & \textbf{0.0784} & \textbf{0.0628} & \textbf{0.1052} & \textbf{0.0755} & \textbf{0.1884} & \textbf{0.3782} & \textbf{0.2785} & \textbf{0.3926}  \\

\midrule
\midrule
\multirow{7}{*}{\makecell{LightGCN}} 
& Normal & 0.2142 & 0.1684 & 0.2894 & 0.1944 & 0.0667 & 0.0535 & 0.0915 & 0.0613 & 0.1559 & 0.3687 & 0.2442 & 0.3804 \\
& T-CE    & 0.2208 & 0.1741 & 0.2975 & 0.2016 & 0.0702 & 0.0558 & 0.0974 & 0.0652 & 0.1625 & 0.3750 & 0.2504 & 0.3833 \\
& DeCA     & 0.2295 & 0.1743 & 0.3026 & 0.2053 & 0.0741 & 0.0614 & 0.1082 & 0.0720 & 0.1620 & 0.3879 & 0.2526 & 0.4005 \\
& BOD    & 0.2335 & 0.1882 & 0.3125 & 0.2168 & 0.0781 & 0.0655 & 0.1102 & 0.0761 & 0.1752 & 0.3915 & 0.2678 & 0.4011 \\
& DCF    & 0.2328 & 0.1877 & 0.3088 & 0.2125 & 0.0754 & 0.0652 & 0.1003 & 0.0748 & 0.1765 & 0.3896 & 0.2711 & 0.3961 \\
& LLaRD  & \underline{0.2393} & \underline{0.1914} & \underline{0.3157} & \underline{0.2169} & \underline{0.0809} & \underline{0.0656} & \underline{0.1192} & \underline{0.0807}  & \underline{0.1903}  & \underline{0.3946}  & \underline{0.2794}  & \underline{0.4077}  \\
\cmidrule{2-14}
& \textbf{ANCHOR} & \textbf{0.2512} & \textbf{0.2025} & \textbf{0.3322} & \textbf{0.2350} & \textbf{0.1084} & \textbf{0.0916} & \textbf{0.1402} & \textbf{0.1035} & \textbf{0.2015} & \textbf{0.4121} & \textbf{0.2866} & \textbf{0.4220} \\

\bottomrule
\end{tabular}%
}
\end{table*}

\subsection{Performance Comparison (RQ1)}
To evaluate the effectiveness and generalizability of our framework, we benchmark our proposed ANCHOR against existing denoising baselines across three datasets and two backbone models. The results are shown in Table~\ref{table:main_all}. The key observations from our experiments are summarized as follows: First, ANCHOR consistently outperforms state-of-the-art denoising baselines across all three datasets and two backbone architectures. This success is attributed to our novel creation-recognition paradigm, which leverages agent-based simulation to generate precise supervisory signals for noise identification and utilizes boundary-adjacent noise to calibrate user preference boundaries. Second, heuristic assumption-based methods generally exhibit limited effectiveness due to their lack of adaptability. These approaches rely on manually crafted rules (e.g., loss thresholds) that fail to generalize across diverse data scenarios or capture complex noise patterns. In contrast, ANCHOR eliminates the dependency on such static assumptions by adopting a creation-recognition paradigm, which autonomously synthesizes adaptive supervisory signals to accurately identify noise without manual intervention. Third, ANCHOR surpasses the second-best baseline, LLaRD, by addressing the lack of explicit supervision in noise identification. While LLaRD effectively leverages LLM-based reasoning to enrich semantic knowledge, it relies on indirect filtering mechanisms (e.g., Information Bottleneck) and fails to explicitly model how noise is generated. ANCHOR overcomes this by actively simulating specific noise behaviors (e.g., misclicks and curiosity) to create direct supervisory signals. This enables the model to learn a more precise decision boundary, effectively distinguishing even hard-to-identify boundary-adjacent noise that LLaRD might overlook.

\begin{table}[t]
\caption{Ablation study of ANCHOR.}
  \centering
  \resizebox{1.0\linewidth}{!}{
    \begin{tabular}{lcccccc}
    \toprule
          & \multicolumn{2}{c}{DBbook2014} & \multicolumn{2}{c}{Book-Crossing} & \multicolumn{2}{c}{MovieLens-1M} \\
    \midrule
          & \multicolumn{1}{c}{R@10} & \multicolumn{1}{c}{N@10} 
          & \multicolumn{1}{c}{R@10} & \multicolumn{1}{c}{N@10} 
          & \multicolumn{1}{c}{R@10} & \multicolumn{1}{c}{N@10} \\
          \cmidrule(r){2-3} \cmidrule(r){4-5} \cmidrule(r){6-7}
    ANCHOR & 0.2512 & 0.2025 & 0.1084 & 0.0916 & 0.2015 & 0.4121  \\
    w/o Gen   & 0.1963  & 0.1511  & 0.0537  & 0.0395  & 0.1438  & 0.3475  \\
    w/o IBR   & 0.2445  & 0.1942  & 0.0834  & 0.0732  & 0.1909  & 0.3961  \\
    w/o Sem   & 0.2323  & 0.1907  & 0.0764  & 0.0648  & 0.1685  & 0.3857  \\
    w/o Rec   & 0.2437  & 0.1914  & 0.0821  & 0.0717  & 0.1936  & 0.4012  \\
    \bottomrule
    \end{tabular}%
  }
  \label{tab:ablation}
\end{table}

\subsection{Ablation Study (RQ2)}
We further conduct an ablation study to examine the contribution of each component in ANCHOR, utilizing LightGCN as the backbone model.
\begin{itemize} [leftmargin=*]
  \item \textbf{w/o Gen}: Replace the out-of-preference noise \textbf{Gen}eration with random negative sampling to construct the supervision signals for the recognizer.
  \item \textbf{w/o IBR}: Exclude the \textbf{I}terative \textbf{B}oundary \textbf{R}efinement mechanism, which means boundary-adjacent noise is not utilized.
  \item \textbf{w/o Sem}: Remove the \textbf{Sem}antic guidance from the noise recognizer, relying exclusively on collaborative ID embeddings to identify noisy interactions while ignoring textual semantic cues.
  \item \textbf{w/o Rec}: Omit the \textbf{Rec}ommendation guidance (i.e., the BPR loss) during recognizer training, optimizing only the binary classification objective without preserving the collaborative preference structure.
\end{itemize}
The performance of all variants is reported in Table~\ref{tab:ablation}, from which several important observations can be drawn. First, removing any component consistently leads to a noticeable performance degradation across all datasets, which demonstrates that each module plays an indispensable role in ANCHOR and jointly contributes to its overall effectiveness. Most notably, the severe drop in w/o Gen demonstrates that explicitly simulating realistic noise provides far superior supervision than heuristic sampling, verifying ANCHOR’s core philosophy. Similarly, the degradation in w/o ABR highlights the importance of boundary-adjacent noise as informative hard negatives for calibrating the decision boundary. Furthermore, w/o Sem shows that semantic cues are essential for identifying subtle noise patterns that ID embeddings alone fail to capture. Finally, w/o Rec reveals the critical role of recommendation-aware guidance; removing the BPR loss decouples noise recognition from the collaborative structure, thereby harming downstream performance.

\begin{figure}[t]
\centering
\includegraphics[width=1.0\linewidth]{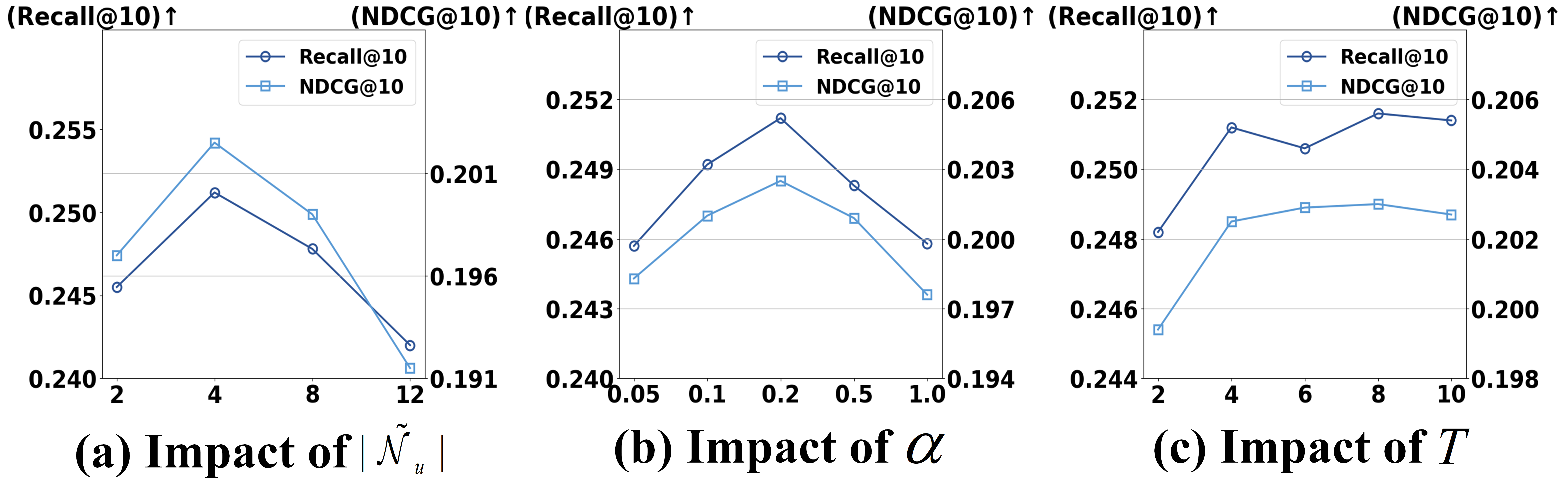}
\caption{Hyperparameter Analysis of ANCHOR.}
\label{fig:hyperparameter}
\end{figure}

\subsection{Hyperparameter Sensitivity (RQ3)}

We analyze ANCHOR's sensitivity to three hyperparameters, as shown in Figure \ref{fig:hyperparameter}. First, when varying the number of generated out-of-preference noise interactions $|\widetilde{\mathcal{N}}_u|$ within $\{2, 4, 8, 12\}$, performance peaks at 4 and subsequently degrades. This suggests that moderate noise aids supervision, whereas excessive noise introduces bias. Second, for the BPR loss weight $\alpha \in \{0.05, 0.1, 0.2, 0.5, 1.0\}$, we observe optimal performance at $\alpha=0.2$. Lower values tend to neglect collaborative signals, while higher values weaken noise recognition. Finally, regarding the adversarial refinement iterations $T \in \{2, 4, 6, 8, 10\}$, performance improves up to $T=4$ and then saturates, indicating that minimal adversarial steps suffice for effective boundary refinement.

\begin{figure}[t]
\centering
\includegraphics[width=1.0\linewidth]{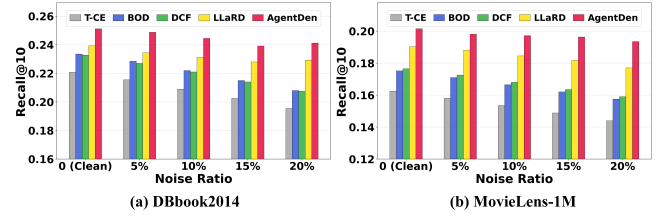}
\caption{Impact comparison \textit{w.r.t.} noise ratio in added interaction data.}
\label{fig:denoise}
\end{figure}
\subsection{Denoising Analysis (RQ4)}
To assess the robustness of ANCHOR under noisy training conditions, we follow common practices in prior work \cite{wang2025unleashing, wang2023efficient} and progressively inject adversarial noisy interactions into the training set at ratios of 5\%, 10\%, 15\%, and 20\%, while keeping the test set unchanged. Figures \ref{fig:denoise} reports the results on the DBbook2014 and MovieLens-1M datasets, respectively. ANCHOR consistently outperforms baselines across all noise levels, demonstrating superior denoising capabilities. Notably, ANCHOR exhibits significantly more stable performance as noise increases. This advantage stems from ANCHOR’s explicit modeling of out-of-preference noise and its adversarial boundary refinement mechanism, which together enable the model to effectively distinguish genuine preference signals from noisy interactions, even in highly noisy environments.

\section{Conclusion}
In this paper, we proposed a Creation-Recognition supervised denoising paradigm and instantiated it with ANCHOR, an agent-based framework for recommendation denoising. The central contribution of ANCHOR is the reformulation of recommendation denoising from heuristic or unsupervised filtering into a supervised learning problem. By proactively simulating behavior-aligned noisy interactions, ANCHOR constructs explicit noise labels and uses them to train a dedicated recognizer, enabling direct and effective noise identification without relying on side information or fragile heuristic assumptions. Moreover, by modeling boundary-adjacent noise through adversarial refinement, ANCHOR exploits informative near-boundary signals to strengthen preference modeling.
\bibliographystyle{ACM-Reference-Format}
\bibliography{reference}

\appendix
\section{Appendix}
\subsection{User Profiling Prompt}
\label{app:user_profiling_prompts}
An example prompt used to initialize the agent with a deep understanding of user preferences based on historical interactions:
\begin{center}
\setlength{\fboxsep}{5pt}
\colorbox{bgcolor}{%
    \begin{varwidth}{\dimexpr0.95\linewidth-2\fboxsep\relax}
    \small
    \emph{Act as a user behavior analyst. You are provided with the interaction history of a specific user to construct a comprehensive behavioral profile. \\
    \textbf{User Interaction History:} \textcolor{magenta}{<InteractionHistory>} \\
    \textbf{Task:}
    1. Analyze the items this user has interacted with to identify core user preferences (e.g., preferred topics, genres, or styles).
    2. Infer behavioral tendencies (e.g., exploratory vs. focused, niche vs. mainstream).
    3. Deduce decision-making patterns (e.g., driven by visual appeal, textual depth, or social trends). \\
    \textbf{Output:} Return a structured list of 5-7 specific insights that capture these dimensions, forming a personalized cognitive blueprint for the user. Do not include negative preferences.}
    \end{varwidth}%
}
\end{center}
\subsection{Out-of-Preference Noise Creation Prompts }
\label{app:noise_creation_prompts}
This appendix provides the detailed prompt templates used by the user simulation agent to create different types of out-of-preference noise.

\vspace{0.5em}

\noindent \textbf{Misclick Noise Prompt.} The prompt for simulating misclick noise instructs the agent to make random selections independent of user preferences:
\begin{center}
\setlength{\fboxsep}{5pt}
\colorbox{bgcolor}{%
    \begin{varwidth}{\dimexpr0.95\linewidth-2\fboxsep\relax}
    \small
    \emph{Act as a user behavior simulator for User \textcolor{olive}{<UserID>}. Given the interaction history \textcolor{magenta}{<History>} and the candidate item set \textcolor{teal}{<CandidateSet>}, simulate accidental interactions. You must select items purely randomly from the candidate list to represent distinct misclicks, inattention, or interface errors. These selections must be stochastic and independent of the user's latent preferences or the item's content. For each generated interaction, output only the selected \textcolor{purple}{<ItemID>} without providing any justification.}
    \end{varwidth}%
}
\end{center}

\vspace{0.5em}

\noindent \textbf{Curiosity-Driven Noise Prompt.} The prompt for curiosity-driven noise guides the agent to select items that trigger temporary exploratory interest:
\begin{center}
\setlength{\fboxsep}{5pt}
\colorbox{bgcolor}{%
    \begin{varwidth}{\dimexpr0.95\linewidth-2\fboxsep\relax}
    \small
    \emph{Act as a user behavior simulator for User \textcolor{olive}{<UserID>}. Analyze the \textcolor{magenta}{<UserProfile>} to understand the user's stable interests. From the candidate item set \textcolor{teal}{<CandidateSet>}, select items that trigger momentary curiosity or a fleeting urge to explore. These items should be distinct from the user's core habits and represent isolated interactions driven by novelty rather than a genuine shift in taste. The objective is to simulate transient exploratory engagements, where the interaction is triggered by novelty but remains incongruent with the user's intrinsic long-term preferences. For each interaction, output the \textcolor{purple}{<ItemID>} and a \textcolor{brown}{<Reason>} identifying the specific element that caused this temporary diversion.}
    \end{varwidth}%
}
\end{center}

\vspace{0.5em}

\noindent \textbf{Caption Bias Noise Prompt.} The prompt for caption bias noise focuses on surface-level textual appeal:
\begin{center}
\setlength{\fboxsep}{5pt}
\colorbox{bgcolor}{%
    \begin{varwidth}{\dimexpr0.95\linewidth-2\fboxsep\relax}
    \small
    \emph{Act as a user behavior simulator for User \textcolor{olive}{<UserID>}. Given the candidate item set \textcolor{teal}{<CandidateSet>} (including titles and descriptions) and the \textcolor{magenta}{<UserProfile>}, simulate the user's susceptibility to textual presentation. Select items where prominent surface-level attributes, including sensational titles, provocative keywords, or intriguing phrasing, exert sufficient allure to elicit an impulsive interaction from this user. Focus on items where the presentation style targets the user's specific attention patterns, even if the content does not align with their long-term goals. For each interaction, output the \textcolor{purple}{<ItemID>} and a \textcolor{brown}{<Reason>} describing the textual element that induced the impulsive click.}
    \end{varwidth}%
}
\end{center}

\vspace{0.5em}

\noindent \textbf{Popularity Bias Noise Prompt.} The prompt for popularity bias noise simulates herd behavior based on item popularity scores:
\begin{center}
\setlength{\fboxsep}{5pt}
\colorbox{bgcolor}{%
    \begin{varwidth}{\dimexpr0.95\linewidth-2\fboxsep\relax}
    \small
    \emph{Act as a user behavior simulator for User \textcolor{olive}{<UserID>}. Given the candidate item set \textcolor{teal}{<CandidateSet>} with \textcolor{cyan}{<PopularityScores>}, and the user's \textcolor{magenta}{<UserProfile>}, simulate herd behavior. Select high-popularity items that this specific user would feel compelled to click due to social proof or trend-following tendencies. Identify cases where the item's widespread hype constitutes a sufficiently strong signal to influence this user's decision, regardless of the mismatch with their personal preferences. For each interaction, output the \textcolor{purple}{<ItemID>} and a \textcolor{brown}{<Reason>} attributing the click to the social influence acting on the user.}
    \end{varwidth}%
}
\end{center}

\vspace{0.5em}

\noindent \textbf{Position Bias Noise Prompt.} The prompt for position bias noise simulates clicks driven by item ranking positions:
\begin{center}
\setlength{\fboxsep}{5pt}
\colorbox{bgcolor}{%
    \begin{varwidth}{\dimexpr0.95\linewidth-2\fboxsep\relax}
    \small
    \emph{Act as a user behavior simulator for User \textcolor{olive}{<UserID>}. Given the \textcolor{teal}{<RankedCandidateList>} where each item has a \textcolor{blue}{<PositionIndex>}, simulate an attention-constrained browsing session for this user. Select items primarily based on their high visibility (low position indices). The selection should reflect a scenario where the user's decision-making is dominated by the convenience of the item's position and limited attention span, effectively bypassing a deep evaluation of the content against the \textcolor{magenta}{<UserProfile>}. For each interaction, output the \textcolor{purple}{<ItemID>} and a \textcolor{brown}{<Reason>} explaining that the interaction was primarily caused by the item's prominent position.}
    \end{varwidth}%
}
\end{center}

\subsection{Adversarial Boundary Refinement Prompts}
\label{app:adversarial_prompts}
The prompt guiding the agent to synthesize ambiguous, near-boundary interactions that challenge the recognizer:
\begin{center}
\setlength{\fboxsep}{5pt}
\colorbox{bgcolor}{%
    \begin{varwidth}{\dimexpr0.95\linewidth-2\fboxsep\relax}
    \small
    \emph{Act as an expert user behavior simulator. Your goal is to generate \textbf{Preference-Boundary Noise}: items that appear highly aligned with the user's preferences on the surface (e.g., similar topics or styles) but fundamentally fail to match their long-term intent or satisfaction. \\
    \textbf{Inputs:}
    \begin{itemize}[leftmargin=*, nosep]
        \item \textbf{User History:} \textcolor{magenta}{<InteractionHistory>}
        \item \textbf{User Preferences:} \textcolor{magenta}{<UserPreferencesSummary>}
        \item \textbf{Adversarial Memory:} \textcolor{orange}{<AccumulatedAdversarialInsights>} (Use this to generate harder samples based on past rounds)
        \item \textbf{Candidate Items:} \textcolor{teal}{<CandidateItems>}
    \end{itemize}
    \textbf{Task:}
    1. Select items from the candidate set that act as "near-miss" counterexamples: strongly related on the surface but poor matches for deep engagement.
    2. Explain the subtle reason why each item lies on the decision boundary (e.g., "misleading title", "right topic but wrong sub-genre").
    3. Leverage the adversarial memory to emphasize patterns that currently fool the noise recognizer. \\
    \textbf{Output:} A structured list of boundary-adjacent noise interactions with hardness tags and reasoning.}
    \end{varwidth}%
}
\end{center}

\vspace{0.5em}

\noindent \textbf{Adversarial Reflection Prompt.} The prompt instructing the agent to analyze recognizer feedback and refine its generation strategy:
\begin{center}
\setlength{\fboxsep}{5pt}
\colorbox{bgcolor}{%
    \begin{varwidth}{\dimexpr0.95\linewidth-2\fboxsep\relax}
    \small
    \emph{Act as an intelligent agent in an adversarial training loop. Your role is to reflect on the discriminator's feedback to produce harder, more informative noise in the next round. \\
    \textbf{Inputs:}
    \begin{itemize}[leftmargin=*, nosep]
        \item \textbf{Adversarial Memory:} \textcolor{orange}{<PastAdversarialMemory>}
        \item \textbf{Discriminator Feedback:} \textcolor{cyan}{<CurrentRoundMetrics>} (Accuracy, specific false positives/negatives, and probability scores)
    \end{itemize}
    \textbf{Task:}
    1. Identify noise patterns that were \textbf{Easy} for the discriminator (correctly detected).
    2. Identify noise patterns that were \textbf{Hard} (successfully deceived the discriminator, causing false negatives).
    3. Analyze the surface similarities or subtle mismatches that contributed to this difficulty.
    4. Formulate updated strategies to generate more deceptive, boundary-aware noise in future rounds. \\
    \textbf{Output:} A reflection summary, identified hard/easy patterns, and concrete strategic guidelines for the next generation cycle.}
    \end{varwidth}%
}
\end{center}
\subsection{Detailed Baseline Introduction}
We use two backbone models:
\begin{itemize} [leftmargin=*]
    \item GMF \cite{he2017neural}: A generalized matrix factorization model that employs a neural architecture to capture user-item interactions.
    \item LightGCN \cite{he2020lightgcn}: A graph-based model that streamlines the GCN architecture by eliminating feature transformation and nonlinear activation.
\end{itemize}
To validate the effectiveness of our proposed approach for implicit feedback denoising, we compare it against five representative denoising baselines, evaluated on the aforementioned recommendation models:
\begin{itemize} [leftmargin=*]
    \item T-CE \cite{wang2021denoising}: A method utilizing truncated cross-entropy to filter out noisy interactions by discarding samples with loss values exceeding a predefined threshold.
    \item DeCA \cite{wang2022learning}: A framework that exploits the prediction discrepancy between two distinct models to identify and mitigate noisy samples.
    \item BOD \cite{wang2023efficient}: An adaptive approach that optimizes sample weights through a bi-level optimization framework to down-weight noisy data.
    \item DCF \cite{he2024double}: A model introducing a double correction mechanism for iterative sample relabeling, simultaneously tackling label noise and data sparsity.
    \item LLaRD \cite{wang2025unleashing}: A framework leveraging LLMs to generate denoising knowledge via graph-based Chain-of-Thought, employing the Information Bottleneck principle to align insights and filter out noise.
\end{itemize}
\end{document}